\documentstyle[prl,aps,psfig]{revtex}
\begin{document}
\draft
\twocolumn[\hsize\textwidth\columnwidth\hsize\csname @twocolumnfalse\endcsname
\title{Flux quantization and superfluid weight in doped antiferromagnets}
\author{Gregory C. Psaltakis}
\address{Department of Physics, University of Crete, and Research Center 
         of Crete, Heraklion, GR-71003, Greece}
\maketitle
\begin{abstract}
Doped antiferromagnets, described by a $t$-$t^{\prime}$-$J$ model and a 
suitable $1/N$ expansion, exhibit a metallic phase-modulated antiferromagnetic 
ground state close to half-filling. Here we demonstrate that the energy of 
latter state is an even periodic function of the external magnetic flux 
threading the square lattice in an Aharonov-Bohm geometry. The period is equal 
to the flux quantum $\Phi_{0}=2\pi\hbar c/q$ entering the Peierls phase factor 
of the hopping matrix elements. Thus flux quantization and a concomitant finite
value of superfluid weight $D_{s}$ occur along with metallic 
antiferromagnetism. We argue that in the context of the present {\em effective}
model, whereby carriers are treated as hard-core bosons, the charge $q$ in the 
associated flux quantum might be set equal to $2e$. Finally, the 
superconducting transition temperature $T_{c}$ is related to $D_{s}$ linearly, 
in accordance to the generic Kosterlitz-Thouless type of transition in a 
two-dimensional system, signaling the coherence of the phase fluctuations of 
the condensate. The calculated dependence of $T_{c}$ on hole concentration
is qualitatively similar to that observed in the high-temperature 
superconducting cuprates.
\end{abstract}
\vskip2pc]

\narrowtext
\section{Introduction}
\label{sec:intro}

During the last few years there has been some evidence that mobile holes in 
doped antiferromagnets, such as the high-$T_{c}$ superconducting copper-oxide 
layers \cite{Bednorz86}, behave much like hard-core bosons. This transmutation 
of statistics, from bare fermionic holes to bosonic vacancy quasiparticles, 
should be understood as an ``emergent phenomenon'' due to the reduced 
dimensionality and the presence of a strongly correlated spin background. 
In the context of the simple fermionic $t$-$J$ model, proposed by Anderson 
\cite{Anderson87-88} to describe such systems, the afore-mentioned evidence 
comes from exact-diagonalization studies of the ground-state energy and the 
static hole-hole correlation function on small clusters 
\cite{Long92-93,Chen94,Eder97}. Indeed, the possibility for a hard-core boson
behavior of the charged vacancies in doped antiferromagnets, opening the way to
Bose-Einstein condensation and the appearance of superconductivity, has been 
suggested by many authors \cite{Kivelson87,Thouless87} in the early days of the
high-$T_{c}$ superconductivity research. Thouless \cite{Thouless87}, in 
particular, argued that due to topological constraints, a vacancy in a 
two-dimensional torus lattice threaded by an external magnetic flux, must be 
transported twice around the ring in order to recover its original 
configuration. Hence flux quantization with an effective charge $q=2e$ may 
result from this period-doubling of the charge $e$ bosons. 

In all the afore-mentioned works, the lack of an effective model for doped 
antiferromagnets expressed in terms of hard-core bosons has prevented the 
systematic study of their flux quantization properties in conjunction with the 
optical and magnetic ones. Such a model, however, has been postulated from the
outset by Psaltakis and Papanicolaou \cite{Psaltakis93} and consists of a 
$t$-$t^{\prime}$-$J$ Hamiltonian and a suitable $1/N$ expansion that provide 
a reasonably simple many-body calculational framework for the study of the 
relevant issues. When leading quantum-fluctuation effects are taken into 
account in the context of this model, the generic experimental features of the 
optical conductivity, the Drude weight, and the total optical weight in the 
cuprates are qualitatively reproduced. In particular, our theory 
\cite{Psaltakis95,Psaltakis96} accounts aptly for the experimentally observed 
$0.5\,\mbox{eV}$ peak of the midinfrared band 
\cite{Cooper90,Orenstein90,Uchida91} and the mass enhancement factor of 
approximately equal to 2 \cite{Orenstein90}. Furthermore, it predicts a finite 
limiting value for the optical conductivity $\sigma(\omega\rightarrow 0)$, 
at finite hole doping, consistent with the residual far-infrared conductivity 
observed in the YBa$_{2}$Cu$_{3}$O$_{6+x}$ family of cuprates \cite{Basov95}. 
Our results are also found to be consistent with relevant exact-diagonalization
data \cite{Dagotto94}.

In view of the quoted evidence from optical experiments in favor of our 
effective model, we undertake in the present paper a systematic study of its 
flux quantization properties in order to provide a more complete assessment 
of the main electromagnetic responses. Our study includes results for the 
superfluid weight $D_{s}$ and the associated superconducting transition 
temperature $T_{c}$. In particular, our explicit numerical estimates for the 
doping dependence of $T_{c}$, including leading quantum-fluctuation effects,
are found to reproduce qualitatively the observed trends in the cuprates
\cite{Uemura89-93,Presland91}.

\section{Effective model}
\label{sec:model}

Our effective model is described by a $t$-$t^{\prime}$-$J$ Hamiltonian 
expressed in terms of Hubbard operators $\chi^{ab}=|a\rangle\langle b|$ as
\begin{equation}
H=-\sum_{i,j}t_{ij}\chi^{0\mu}_{i}\chi^{\mu 0}_{j}
+{\textstyle\frac{1}{2}}J\sum_{\langle i,j\rangle}
(\chi^{\mu\nu}_{i}\chi^{\nu\mu}_{j}-\chi^{\mu\mu}_{i}\chi^{\nu\nu}_{j}) \;,
\label{eq:Hamiltonian}
\end{equation}
where the index 0 corresponds to a hole, the Greek indices $\mu,\nu,\ldots$ 
assume two distinct values, for a spin-up and a spin-down electron, and the 
summation convention is invoked. Here $J$ is the antiferromagnetic 
spin-exchange interaction between nearest-neighbor sites ${\langle i,j\rangle}$
on a square lattice endorsed with periodic boundary conditions and a total
number of sites $\Lambda=\Lambda_{x}\times\Lambda_{y}$, where 
$\Lambda_{x}=\Lambda_{y}$. For the hopping matrix elements $t_{ij}$ we assume
\begin{equation}
t_{ij}=\left\{\begin{array}{cl}
       t & \mbox{if $i,j$ are nearest neighbors} \\
       -t^{\prime} & \mbox{if $i,j$ are next nearest neighbors} \\
       0 & \mbox{otherwise} \;.
       \end{array}\right.
\label{eq:hopping}
\end{equation}
The conventions in (\ref{eq:hopping}) incorporate opposite signs for $t$ and 
$t^{\prime}$ as dictated by quantum-chemistry calculations 
\cite{Hybertsen90,Sawatzky90} for Cu-O clusters and fits of the shape of the 
Fermi surface observed by angle-resolved photoemission spectroscopy \cite{Yu91}.
In Ref.~\onlinecite{Psaltakis93} we generalized the local constraint associated
with (\ref{eq:Hamiltonian}) to $\chi^{00}_{i}+\chi^{\mu\mu}_{i}=N$, where $N$ 
is an arbitrary integer, and considered the commutation properties of the 
$\chi^{ab}$'s to be those of the generators of the U(3) algebra. A generalized 
Holstein-Primakoff realization for the latter algebra reads
\begin{eqnarray}
\chi^{00}_{i} & = & N-\xi^{\mu\ast}_{i}\xi^{\mu}_{i}, \;\;\;
\chi^{\mu\nu}_{i}=\xi^{\mu\ast}_{i}\xi^{\nu}_{i} \;, \nonumber \\
& & \label{eq:repres} \\
\chi^{0\mu}_{i} & = & (N-\xi^{\nu\ast}_{i}\xi^{\nu}_{i})^{1/2}\xi^{\mu}_{i},
\;\; \chi^{\mu 0}_{i}=\xi^{\mu\ast}_{i}(N-\xi^{\nu\ast}_{i}\xi^{\nu}_{i})^{1/2},
\nonumber
\end{eqnarray}
where the $\xi^{\mu}_{i}$ are Bose operators,
$[\xi^{\mu}_{i},\xi^{\nu\ast}_{j}]=\delta_{ij}\delta^{\mu\nu}$. Note that
the local constraint, giving rise to the hard-core character of the bosons, has
been explicitly resolved in (\ref{eq:repres}). One can then develop a 
perturbation theory based on the $1/N$ expansion, restoring the relevant 
physical value $N=1$ at the end of the calculation. 

In the presence of an external magnetic flux $\Phi$, threading the 
two-dimensional lattice in an Aharonov-Bohm torus geometry, the hopping matrix 
elements $t_{ij}$ are modified by the well-known Peierls phase factor 
and should be substituted in (\ref{eq:Hamiltonian}) according to
\begin{equation}
t_{ij}\leadsto t_{ij}e^{iA_{ij}} \;, \;\;\;
\mbox{with} \;\; A_{ij}=\frac{2\pi\Phi}{\Lambda_{x}\Phi_{0}}
({\bf R}_{i}-{\bf R}_{j})\cdot{\bf e}_{x} \;.
\label{eq:Peierls}
\end{equation}
Here ${\bf R}_{i}$ is the position vector for site $i$, ${\bf e}_{x}$ is the 
unit vector along the $x$-axis encircling the flux lines, and 
$\Phi_{0}=2\pi\hbar c/q$ is the so-called flux quantum. Conventionally, the 
charge $q$ of the carriers entering $\Phi_{0}$ is, of course, equal to the 
electronic charge $e$. However, the arguments of Thouless \cite{Thouless87} 
quoted in the Introduction imply that a vacancy actually ``feels'' twice as 
much external flux. In the context of the present {\em effective} model this 
may be accounted for by an extra factor of two in the expression 
(\ref{eq:Peierls}) for the $A_{ij}$ which can be readily absorbed in a 
redefinition of $q$ as $q=2e$. Evidently, this reasoning does not constitute a 
rigorous justification for the assignment $q=2e$ in the flux quantum 
$\Phi_{0}$. The latter justification can be provided only by an ab initio 
derivation of an effective Hamiltonian for the hard-core boson vacancies, 
starting from a realistic electronic model for the cuprates. At present such 
a program is out of reach. Hence this work will be content with the study of 
the flux quantization properties of the effective model described by 
(\ref{eq:Hamiltonian})--(\ref{eq:Peierls}), given the flux quantum constant 
$\Phi_{0}$.

In the large-$N$ limit ``condensation'' occurs, i.e., the Bose operators 
$\xi^{\mu}_{i}$, $\mu=1,2$, become classical commuting fields. For uniform
density states these complex number amplitudes may then be parametrized as
\begin{eqnarray}
\xi^{1}_{i} & = & \sqrt{Nn_{e}}\,\cos\left(\frac{\theta_{i}}{2}\right)
e^{i\psi_{i}/2}e^{-i\phi_{i}/2} \;, 
\nonumber \\
& & \label{eq:param} \\
\xi^{2}_{i} & = & \sqrt{Nn_{e}}\,\sin\left(\frac{\theta_{i}}{2}\right)
e^{i\psi_{i}/2}e^{i\phi_{i}/2} \;,
\nonumber
\end{eqnarray} 
where $n_{e}$ is the average electronic density, the angles $\theta_{i}$ and
$\phi_{i}$ determine the local spin direction, while the remaining parameter 
$\psi_{i}$ determines the local phase of the condensate. As shown in 
Ref.~\onlinecite{Psaltakis93}, close to half-filling ($n_{e}\lesssim 1$) and
for a sufficiently large $t^{\prime}$, the ground state of 
(\ref{eq:Hamiltonian}) is described by a planar spin configuration 
($\theta_{i}=\pi/2$) in which the local twist angles {\em and} phases are 
modulated according to 
\begin{equation}
\phi_{i}={\bf Q}\cdot{\bf R}_{i} \;, \;\;\;
\psi_{i}={\bf Q}^{\prime}\cdot{\bf R}_{i} \;,
\label{eq:ansatz}
\end{equation}
where ${\bf Q}=(\pi,\pi)$ is the usual spin-modulating antiferromagnetic 
wavevector and ${\bf Q}^{\prime}=(\pi,-\pi)$ is an unusual phase-modulating 
wavevector. We should note here that the excitation spectrum above this ground 
state is gapless \cite{Psaltakis93}, hence, as quoted in the Introduction,
the limiting value of the optical conductivity $\sigma(\omega\rightarrow 0)$ 
remains finite, at finite hole doping. However, as the half-filled-band limit 
is approached ($n_{e}\rightarrow 1$) the quantity 
$\sigma(\omega\rightarrow 0)/(1-n_{e})$ becomes increasingly depressed 
\cite{Psaltakis95}. This trend is consistent with the ubiquitous ``pseudogap'' 
behavior observed in the optical and magnetic properties of underdoped cuprates
\cite{Coleman-Norman98} and provides further support to the relevance of the 
metallic phase-modulated antiferromagnetic (AF) ground state under 
consideration. The question that now poses is how this ground state will 
respond to the presence of an external magnetic flux $\Phi$?

\section{Flux quantization and superfluid weight}

\begin{figure}[h]
\centerline{\psfig{figure=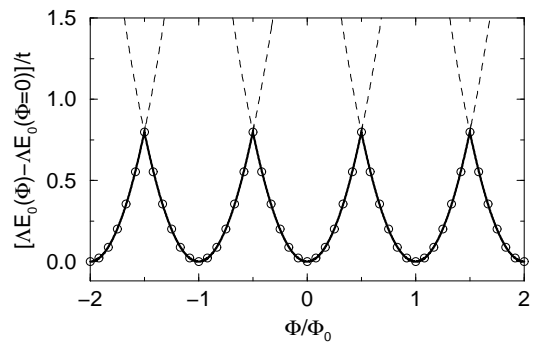,width=7.0cm}}
\caption{\label{fig:AF-flux} Ground-state energy vs external magnetic flux, for
$\varepsilon=0.45$, $t/J=1.0$, and $1-n_{e}=0.10$. The zero-flux energy is 
subtracted off as to normalize the values. Solid line: the analytic result in 
the infinite lattice limit ($\Lambda\rightarrow\infty$), according to 
Eqs.~(\protect\ref{eq:energy-diff})--(\protect\ref{eq:envelope}). Dashed lines:
remnants of the crossing energy-level parabolas discussed in the text. Open 
circles: numerical minimization results for the ground-state energy on a finite
lattice ($\Lambda=20\times 20$), as determined by 
Eqs.~(\protect\ref{eq:total-energy})--(\protect\ref{eq:kinetic-exchange}). 
Evidently, the finite lattice numerical data (open circles) confirm the infinite
lattice limit analytic result (solid line).}
\end{figure}

Following an argument by Yang \cite{Yang62} we note that, in the presence of 
$\Phi$, the reciprocal lattice is displaced from the origin by 
$2\pi\Phi/(\Lambda_{x}\Phi_{0})$ along the $x$-axis. The quantization of flux 
therefore depends on whether the ground-state energy of the system changes 
under this momentum boost. Given that the spin-exchange part of the Hamiltonian
(\ref{eq:Hamiltonian})--(\ref{eq:Peierls}) does not couple directly to the
magnetic flux it is plausible that, at least in the large-$N$ limit, the 
condensate will respond in such a way as to leave its spin-modulating 
wavevector ${\bf Q}$ intact and simply adjust its phase-modulating wavevector 
${\bf Q}^{\prime}$ to a new value. In other words, we anticipate that in this 
classical (large-$N$) limit, the rigidity of the ground state against the 
intrusion of the external magnetic flux comes solely from the phase 
fluctuations of the condensate. These heuristic arguments lead us to consider 
the ansatz (\ref{eq:ansatz}) with the following modulating wavevectors 
\begin{equation}
{\bf Q}=(\pi,\pi)  \;, \;\;\;
{\bf Q}^{\prime}=(\pi,-\pi)-\left(\frac{4\pi m}{\Lambda_{x}},0\right) \;,
\label{eq:wavevectors}
\end{equation}
where $m$ is an arbitrary integer. 
Inserting (\ref{eq:wavevectors}) into (\ref{eq:param})--(\ref{eq:ansatz}), the 
Hamiltonian (\ref{eq:Hamiltonian})--(\ref{eq:Peierls}) takes the form 
$H(\Phi)=N^{2}\Lambda E_{0}(\Phi)$, where $E_{0}(\Phi)$ is the classical energy
per lattice site for the value of physical interest $N=1$. More explicitly,
taking carefully the infinite lattice limit ($\Lambda\rightarrow\infty$) we 
have that
\begin{equation}
\Lambda E_{0}(\Phi)-\Lambda E_{0}(\Phi=0)=8t^{\prime}\pi^{2}n_{e}(1-n_{e})
\left(\frac{\Phi}{\Phi_{0}}-m\right)^{2} \;.
\label{eq:energy-diff}
\end{equation}
Thus for each integer $m$ we get an individual many-body energy level that 
depends quadratically on $\Phi$. The ground-state energy is given by the lower 
envelope of these crossing energy-level parabolas and is characterized 
analytically by the condition
\begin{equation}
\left|\frac{\Phi}{\Phi_{0}}-m\right|\leq\frac{1}{2} \;, 
\;\;\; \mbox{with} \;\; m=0,\pm1,\pm2,\ldots \;.
\label{eq:envelope}
\end{equation}
In Fig.~\ref{fig:AF-flux} we depict by solid line the ground-state energy 
calculated according to (\ref{eq:energy-diff})--(\ref{eq:envelope}), for typical
values of the parameters $\varepsilon=t^{\prime}/t$, $t/J$, and the hole 
concentration $(1-n_{e})$. We also depict by dashed lines the remnants 
of the individual crossing energy levels (\ref{eq:energy-diff}). Evidently,
the ground-state energy (solid line) is an even periodic function of the 
external magnetic flux $\Phi$, with a macroscopic energy barrier between 
different flux minima, in accordance with the Byers and Young \cite{Byers61} 
characterization of a superconductor. The period is equal to $\Phi_{0}$ and
therefore the assignment $q=2e$, discussed earlier on, leads to agreement with 
the observed flux quantization in the high-$T_{c}$ superconducting copper-oxide
layers \cite{Gough87}. In order to establish firmly the analytic result 
(\ref{eq:energy-diff})--(\ref{eq:envelope}), and thus the heuristic arguments
involved in (\ref{eq:ansatz})--(\ref{eq:wavevectors}), we have also minimized 
numerically the classical energy $\Lambda E_{0}(\Phi)$ obtained 
by inserting directly (\ref{eq:param}) into 
(\ref{eq:Hamiltonian})--(\ref{eq:Peierls}),
\begin{equation}
\Lambda E_{0}(\Phi)={\cal E}_{1}+{\cal E}_{2} \;,
\label{eq:total-energy}
\end{equation}
where
\begin{eqnarray}
{\cal E}_{1} &=& -n_{e}(1-n_{e})\sum_{i,j}t_{ij} 
\nonumber \\
&& \left[\cos\frac{\theta_{i}}{2}\cos\frac{\theta_{j}}{2} 
\cos\left(A_{ij}+\frac{\psi_{i}-\psi_{j}-\phi_{i}+\phi_{j}}{2}\right)\right.
\nonumber \\
&& \left.+\sin\frac{\theta_{i}}{2}\sin\frac{\theta_{j}}{2} 
\cos\left(A_{ij}+\frac{\psi_{i}-\psi_{j}+\phi_{i}-\phi_{j}}{2}\right)\right] \;,
\nonumber \\
&& \label{eq:kinetic-exchange} \\
{\cal E}_{2} &=& \frac{n_{e}^{2}}{4}J\sum_{\langle i,j\rangle}
[\cos\theta_{i}\cos\theta_{j}
+\sin\theta_{i}\sin\theta_{j}\cos(\phi_{i}-\phi_{j})-1] \;.
\nonumber
\end{eqnarray}
The minimization of (\ref{eq:total-energy})--(\ref{eq:kinetic-exchange}) was 
carried out by a relaxation method. Excellent agreement with the analytic 
result (\ref{eq:energy-diff})--(\ref{eq:envelope}) was obtained already for 
lattices with $\Lambda=20\times 20$, and for all choices of the parameters
$\varepsilon$, $t/J$, and $n_{e}$, within the range of stability of the 
phase-modulated AF ground state. A specific example of this agreement is 
evidenced in Fig.~\ref{fig:AF-flux}, where the open circles correspond to the 
numerical minimization data.

Let us now turn our attention to the superfluid weight (or helicity modulus) 
$D_{s}$ given by the curvature of the infinite lattice limit of the 
ground-state energy $\Lambda E(\Phi)$ at $\Phi=0$ 
\cite{Byers61,Yang62,Scalapino92-94}, 
\begin{equation}
D_{s}=\Lambda\left(\frac{\Phi_{0}}{2\pi}\right)^{2}
\left[\frac{\partial^{2}E(\Phi)}{\partial\Phi^{2}}\right]_{\Phi=0} \;.
\label{eq:D_s:definition}
\end{equation}
$D_{s}$ determines the ratio of the density of the superfluid charge carriers 
to their mass, and is related to the directly measurable in-plane London 
penetration depth $\lambda_{L}$ by $D_{s}=c^{2}/(4\pi e^{2}\lambda_{L}^{2})$.
Quite generally, $E(\Phi)$ has an $1/N$ expansion of the form
$E(\Phi)=N^{2}E_{0}(\Phi)+NE_{1}(\Phi)+\cdots\;$ which leads via 
(\ref{eq:D_s:definition}) to a corresponding expansion for the superfluid 
weight $D_{s}=N^{2}D_{s}^{(0)}+ND_{s}^{(1)}+\cdots\;$. Hence by exploiting the 
large-$N$ limit result (\ref{eq:energy-diff})--(\ref{eq:envelope}) we get 
immediately the expression for the leading term $D_{s}^{(0)}$,
\begin{equation}
D_{s}^{(0)}=4t^{\prime}n_{e}(1-n_{e}) \;.
\label{eq:Ds(0)}
\end{equation}
Our earlier arguments show that $D_{s}^{(0)}$ is a measure of 
the stiffness of the classical phase fluctuations of the condensate. 
Furthermore, (\ref{eq:Ds(0)}) implies $D_{s}^{(0)}=D_{0}$, where $D_{0}$ is the
leading term in the $1/N$ expansion of the Drude weight 
$D=N^{2}D_{0}+ND_{1}+\cdots\;$, studied in Ref.~\onlinecite{Psaltakis96} using 
Kubo formalism for the current-current correlations. We have also verified, 
by a straightforward but lengthy calculation of $E_{1}(\Phi)$ and the use of 
(\ref{eq:D_s:definition}), that $D_{s}^{(1)}=D_{1}$. Due to the analytic 
structure of the $1/N$ expansion, these results signify the term-by-term 
validity of the identity $D_{s}=D$. Strictly speaking, of course, we have 
checked explicitly that $D_{s}=D$ only up to and including terms 
$D_{s}^{(1)}=D_{1}$, i.e., only up to and including leading quantum-fluctuation 
effects \cite{comment}. This, however, is sufficient for most practical 
purposes and permits us to exploit our calculations of the Drude 
weight, in the present study of the superfluid weight. For instance, the weight
$D_{s}=D$ including leading quantum-fluctuation effects, is found 
\cite{Psaltakis96} to increase linearly with small hole concentration 
$(1-n_{e})$ away from the half-filled-band limit ($n_{e}=1$). This trend, 
present already in (\ref{eq:Ds(0)}), is a fundamental characteristic of doped 
antiferromagnets. At higher doping values $D_{s}=D$ eventually saturate and 
then start to decrease. Note that the vanishing overlap between the opposite 
sublattice spin states, along with the absence of quantum fluctuations in the
large-$N$ limit, leaves the direct hopping $t^{\prime}$ between same sublattice
sites as the only relevant process of charge transport in this classical 
approximation. This argument makes plausible the independence of $D_{s}^{(0)}$
from $t$ and $J$ seen in (\ref{eq:Ds(0)}). However, the leading 
quantum-fluctuation correction $D_{s}^{(1)}=D_{1}$ involves already a 
non-trivial dependence on the latter couplings. 

It should be noted that when $t^{\prime}=0$, the present model reduces to the 
simple $t$-$J$ model where in the physically relevant regime, i.e., close to 
half-filling, the uniform density state under study becomes unstable against 
phase separation into an insulating (hole-poor) antiferromagnetic region and a 
conventional metallic (hole-rich) ferromagnetic region \cite{Marder90}. In the 
latter phase separated state no flux quantization and finite superfluid weight
occurs. A vanishing superfluid weight has been suggested also by the Quantum 
Monte Carlo studies of the simple Hubbard model \cite{Scalapino92-94}, although 
the corresponding exact-diagonalization studies of the fermionic $t$-$J$ model 
\cite{Dagotto94} are not conclusive close to half-filling, due to the very 
small lattice sizes (e.g., $4\times 4$) used. Indeed, the finite-size effects 
in the numerical studies of the latter system are particularly large because of
the presence of phase separation in the ground state \cite{Hellberg97}. Our 
observations here underline the importance of the next-nearest-neighbor hopping
$t^{\prime}$ to the ability of the mobile holes in sustaining a uniform density
state that displays flux quantization and a finite superfluid weight. In this 
respect it is useful to remind that the effective hopping parameter 
$t^{\prime}$ accounts for the large oxygen-oxygen overlap integrals present in 
the original CuO$_{2}$ planes \cite{Hybertsen90,Sawatzky90,Yu91}.

We will complete our report with a discussion of the expected transition 
temperature to the charged superfluid, i.e., superconducting, state under 
study. At a finite temperature $T$, the ratio of the thermal de Broglie 
wavelength of the charge carriers to their average distance is proportional to 
$\sqrt{D_{s}/(k_{B}T)}$, where $D_{s}$ is the zero-temperature value determined
by (\ref{eq:D_s:definition}). Hence a naive application of the criterion for 
the occurrence of Bose-Einstein condensation in an ideal boson gas, whereby the
latter ratio should become of order unity, suggests a transition temperature 
$T_{c}$ of the form
\begin{equation}
k_{B}T_{c}=AD_{s} \;,
\label{eq:Tc}
\end{equation}
where $A$ is a dimensionless constant of order unity. Of course, in the strictly
two-dimensional model of continuous symmetry under study, a bona fide finite 
temperature phase transition can only be of the Kosterlitz-Thouless type which, 
nevertheless, leads again to an expression of the form (\ref{eq:Tc}). 
Indeed, the $\psi_{i}$-structure of the classical Hamiltonian 
(\ref{eq:total-energy})--(\ref{eq:kinetic-exchange}) is a generalization of the
two-dimensional $XY$ model where the latter transition is well studied. In this
context, it is important to note that a ``universal'' linear relation of the 
form (\ref{eq:Tc}) has been established experimentally in the cuprates by 
Uemura {\it et al}. \cite{Uemura89-93} in their remarkable study of $T_{c}$ as
a function of the zero-temperature value of $\lambda_{L}^{-2}\propto D_{s}$. In
the large-$N$ limit, the $D_{s}$ appearing in (\ref{eq:Tc}) is just equal to
$D_{s}^{(0)}$ and the corresponding critical temperature $T_{c}^{(0)}$ should 
be interpreted as the ordering temperature for the classical phase fluctuations
of the condensate, in analogy with the analysis of Emery and Kivelson 
\cite{Emery95} of the classical phase fluctuations of the conventional BCS 
order parameter. The higher order terms in the $1/N$ expansion of $D_{s}=D$ 
capture the effects of the quantum fluctuations and renormalize downwards these
weights \cite{Psaltakis96}, thereby reducing the corresponding value of $T_{c}$.

\begin{figure}[h]
\centerline{\psfig{figure=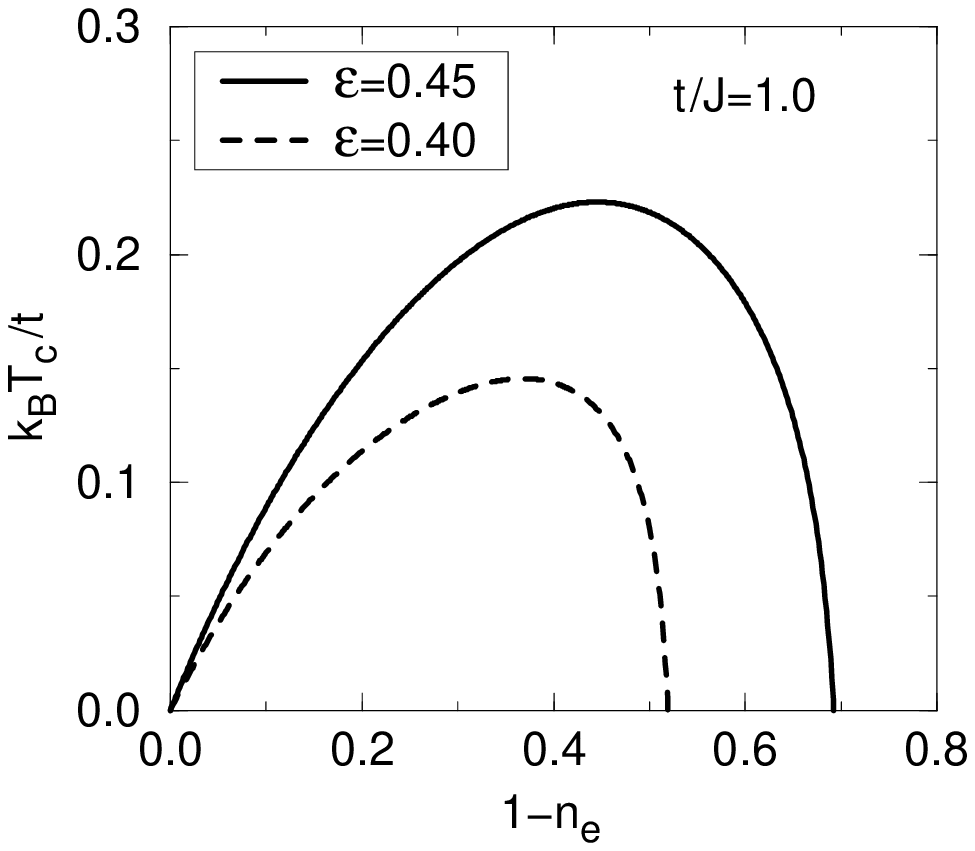,width=5.5cm}}
\caption{\label{fig:hightc} Superconducting transition temperature vs hole 
concentration, for $t/J=1.0$ and $\varepsilon=0.45$ (solid line) or 
$\varepsilon=0.40$ (dashed line), according to Eq.~(\protect\ref{eq:Tc}) 
with the inclusion of leading quantum-fluctuation effects.}
\end{figure}

Following the prescription of Emery and Kivelson \cite{Emery95}, we have 
applied (\ref{eq:Tc}) with $A=0.9$; a numerical value extracted from the 
two-dimensional $XY$ model \cite{Gupta88}. Using the calculated $D_{s}=D$ 
of Ref.~\onlinecite{Psaltakis96}, with the inclusion of the leading 
quantum-fluctuation correction $D_{s}^{(1)}=D_{1}$, we depict in 
Fig.~\ref{fig:hightc} the superconducting transition temperature $T_{c}$ as a 
function of the hole concentration $(1-n_{e})$. Evidently, the dependence of 
$T_{c}$ on $(1-n_{e})$ reflects that of $D_{s}$ and reproduces qualitatively 
the observed trends in the cuprates \cite{Uemura89-93,Presland91}. With an 
estimated $J/k_{B}\approx 1500\,\mbox{K}$ in the latter materials 
\cite{Shamoto93}, the value of $T_{c}$ at optimum doping $(1-n_{e})=0.44$ 
($0.36$), seen in the solid (dashed) line of Fig.~\ref{fig:hightc}, 
is $T_{c}\approx 335\,\mbox{K}$ ($218\,\mbox{K}$). This predicted value of 
$T_{c}$, signaling the coherence of the phase fluctuations of the condensate, 
should be regarded as an upper bound to an actual transition temperature 
because of the neglect of impurity disorder, higher-order quantum fluctuations,
etc. From Fig.~\ref{fig:hightc} we also note that with further hole doping 
$T_{c}$ starts to decrease while beyond a critical doping value it vanishes, 
as the phase-modulated AF configuration, around which the present $1/N$ 
expansion is carried out, becomes unstable.

\section{Conclusions}

In this paper, we have demonstrated that flux quantization and a 
concomitant finite value of superfluid weight $D_{s}$ occur in the metallic 
phase-modulated AF ground state of the $t$-$t^{\prime}$-$J$ model 
(\ref{eq:Hamiltonian}). The classical phase fluctuations of the condensate are 
shown to control the leading term in the $1/N$ expansion of $D_{s}$. 
By appealing to the universality class of the two-dimensional $XY$ model, 
the corresponding superconducting transition temperature $T_{c}$ is related to 
$D_{s}$ linearly, via (\ref{eq:Tc}). The inclusion of leading 
quantum-fluctuation effects in $D_{s}$ provides then a reasonable estimate for 
the order of magnitude and the doping dependence of $T_{c}$ in the cuprates. 
The latter dependence is of particular importance as it emerges from a 
consistent many-body $1/N$ expansion that preserves, at each order of 
perturbation theory, the local constraint, implied by the strong-correlation 
effects. These results support our effective description of the charge carriers
in terms of hard-core bosons.

\section*{Acknowledgments}

It is a pleasure to thank X. Zotos, E. Manousakis, and G. Varelogiannis for 
stimulating discussions. This work was supported by Grant No. 
$\Pi$ENE$\Delta$95-145 from the Greek Secretariat for Research and Technology.

\begin{thebibliography}{99}
\bibitem{comment} Note that for a conventional BCS superconductor with a finite
energy gap, a case radically different from the one under study, the identity 
$D_{s}=D$ is proved in Ref.~\onlinecite{Scalapino92-94}.
\end{thebibliography}
\end{document}